# Interplay between the magnetic and electric degrees-of-freedom in multiferroic $Co_3TeO_6$

Wen-Hsien Li[1,*], Chin-Wei Wang[1], Daniel Hsu[1], Chi-Hung Lee[1], Chun-Ming Wu[1], Chih-Chieh Chou[2], Hung-Duen Yang[2], Yang Zhao[3], Sung Chang[3], Jeffrey W. Lynn[3], and Helmuth Berger[4]

[1]*Department of Physics and Center for Neutron Beam Applications, National Central University, Jhongli 32001, Taiwan*

[2]*Department of Physics and Center for Nanoscience and Nanotechnology, National Sun Yat-Sen University, Kaohsiung 80424, Taiwan*

[3]*NIST Center for Neutron Research, National Institute of Standards and Technology, Gaithersburg, Maryland 20899, USA*

[4]*Institute of Physics of Complex Matter, EPFL, Lausanne, Switzerland*

Abstract

Neutron diffraction, magnetic susceptibility, specific heat, and dielectric constant measurements of single crystal $Co_3TeO_6$ have been measured to study the interplay between the ferroelectricity and magnetic order. Long range incommensurate magnetic order develops below $T_{M1}$=26 K, which is followed by three additional zero-field phase transitions at $T_{M2}$=19.5 K, $T_{M3}$=18 K, and $T_{M4}$=16 K where the incommensurate order changes and commensurate order develops. In magnetic fields up to 14 T we find that the magnetic intensities and incommensurate wave vector are dramatically altered as ferroelectricity develops, with a fifth abrupt transition around 10 T. The overall behavior characterizes $Co_3TeO_6$ as a type-II multiferroic.

*Corresponding author: Tel.: 886-3-4227151 ext. 65317; Fax: 886-3-2807404; E-mail: whli@phy.ncu.edu.tw

Ferroelectricity is a physical phenomenon where spontaneous electrical polarization **P** appears below a certain temperature $T_C$. Ferroelectric order can arise in a number of ways, such as from a structural change, a spin-exchange interaction, or spin-orbit interaction that triggers relative displacements of the positive and negative ions.[1-6] Multiferroics, where both ferroelectric and magnetic order coexist, are quite uncommon, but are of particular interest both to understand the fundamental interactions between the two types of order as well as for the potential for practical applications.[7-9] For conventional (type-I, or proper) ferroelectrics the two types of order must be associated with different atoms, with generally a weak interaction between the two order parameters. For type-II (improper) ferroelectrics, on the other hand, the ferroelectric displacements originate from some other type of ordering, and of particular interest here is when this is magnetic order.[10-17] Understanding the nature of this coupling is a primary motivation for investigating these materials both for fundamental purposes and to enable the tailoring of materials properties and designing devices with the necessary magnetoelectric (ME) coupling for spintronics applications. It is generally the case that the strength of the ME effect of type-II multiferroics, where the order parameters of **P** and **M** are closely coupled, is stronger than that of type-I multiferroics, where **P** and **M** are essentially decoupled to a first approximation.[8,9] In this context, the novel metal tellurates $M_3TeO_6$, where $M$ is a first-row transition metal, have been shown to be rich in crystalline chemistry.[18-22] Ferroelectricity,[20-22] ferromagnetic and antiferromagnetic spin orders,[18,19,23] complex incommensurate spin structures,[20,21] and magnetic-field-driven polarization[22] have all been observed. Here, we report on the observations of a strong interplay between the order parameters of ferroelectricity and both commensurate and incommensurate magnetic order in single crystal cobalt tellurate $Co_3TeO_6$. We find $Co_3TeO_6$ to be a type-II multiferroic.

Single crystals of $Co_3TeO_6$ were synthesized via chemical vapor transport redox reactions.[24] The $Co_3O_4$, $TeO_2$, and $CoCl_2$ powders were mixed thoroughly using a molar ratio of 4:3:1 before being loaded into one end of a silica tube. The tube was then evacuated to $10^{-5}$ torr, and filled with HCl gas to act as the transporting agent, before being sealed off. The ampoule was subsequently placed in a two-zone furnace, with the temperatures of the charge and growth zones set to 973 and 873 K, respectively. This allowed the transportation of the starting materials from the charge zone to epitaxially grow into single crystals in the growth zone. The resultant single crystals were dark-violet in color. The single crystal used in the present measurements weighed 101 mg, with a size of $14.1\times2.2\times0.9$ $mm^3$. A number of small crystals also were crushed into powder for magnetic measurements, and x-ray and neutron powder diffraction measurements. The magnetic susceptibility and specific heat measurements were performed on a Quantum Design Physical Properties Measurement System (PPMS). The dielectric measurements were made using the Agilent E4980A precision LCR Meter, with an *ac* voltage of 1 V applied alone the *a*-axis. The neutron diffraction measurements were conducted on BT-7 and BT-9 triple-axis spectrometers at the NIST Center for Neutron Research using pyrolytic graphite PG(002) monochromator crystals to select an incident wavelength of $\lambda$=2.359 Å, with PG filter to suppress higher-order wavelength contaminations. Single detector measurements were made with a PG(002) energy analyzer, while a position sensitive detector (PSD) was employed in diffraction mode (no analyzer) on BT-7 to map regions of reciprocal space.[25] The magnetic field for the neutron diffraction experiments was provided by a 15 T OXFORD superconducting magnet. The single crystal was mounted to allow access to the *a-b* scattering plane. The field dependent measurements utilized the same scattering plane, with the magnetic field being applied along the *c*-axis. Uncertainties where indicated represent one standard deviation.

$Co_3TeO_6$ crystallizes into a monoclinic symmetry with a space group of C2/c. At room temperature the lattice parameters are *a*=14.7526(2) Å, *b*=8.8139(1) Å, *c*=10.3117(1) Å, and β=94.905(2)°, which then are used to define the reciprocal lattice constants and reciprocal lattice units (r.l.u) (*h,k,l*). The Te and Co atoms can be viewed as being arranged in the crystallographic *b*-*c* planes, with *a* being the longest axis.[24] There are two Te sites and five Co sites in the chemical unit cell that are crystallographically distinguishable, with a total of 36 Co spins. The Co ions form a significantly distorted hexagonal arrangement and the Te ions are located in the channels. The Co-O bond lengths for the $CoO_6$ octahedra and $CoO_4$ tetrahedra vary widely; the differences can be as high as 50%,[24] which result in widely varying magnetic interactions that compete within the unit cell. At 40 K, where no significant electrical or magnetic activity[22] has been found, the crystalline symmetry of $Co_3TeO_6$ remains monoclinic C2/c.

Upon cooling from room temperature, the thermal profile of the magnetic susceptibility χ can be described by Curie-Weiss behavior down to ≈45 K, where χ(T) departs from the Curie-Weiss profile, indicating the development of magnetic correlations. A downturn in χ(T) appears at 26 K, with an abrupt drop at 18.5 K, as shown in Fig. 1. Interestingly, an applied magnetic field **B** dramatically alters the thermal profile of χ at lower temperatures, with the temperatures at which these anomalies appear noticeably shifting with **B**. There are two readily identifiable anomalies in the χ(T) curves, designated as $T_{M1}$ and $T_{M3}$. Both $T_{M1}$ and $T_{M3}$ shift to considerably lower temperatures with increasing **B**.

To investigate the nature of the magnetic order, single crystal neutron diffraction maps in the [*hk*0] plane are shown in Fig. 2. At 22 K (Fig. 2(a,b)) we find that four incommensurate (ICM) peaks that surround some of the integer nuclear Bragg peak positions have appeared, demonstrating that incommensurate magnetic order has developed with wave vector $\vec{q}$=(0.355, 0.155, 0). In addition, around the strongest (0, 2, 0) position we see clear diffuse scattering originating from spin excitations. At 16.5 K (Fig. 2(c)) a phase transition clearly has occurred where the k component of the incommensurability has vanished. Upon further cooling to low temperature (Fig. 2(d)) the k component of the incommensurability has been restored, indicating that yet another magnetic phase transition has occurred. Figures 2(e-g) present maps of the temperature dependence of the intensities and positions of these incommensurate peaks, clearly showing that the modulation vector shifts dramatically with temperature.[27]

Fig. 3 presents quantitative values for the temperature dependence of the peak intensities and positions, together with the heat capacity and dielectric constant, to facilitate a direct comparison of the temperature evolution of the various measured properties. Overall four phase transitions can be readily identified, beginning with $T_{M1}$=26 K which marks the onset of long range incommensurate magnetic order. Fig. 3(a) shows that there is no change in the structural Bragg peaks below $T_{M1}$, which is confirmed by neutron powder diffraction patterns taken at 30 and 22 K. Thus initially the magnetic structure is purely incommensurate. However, below $T_{M2}$=19.5 K the commensurate Bragg peaks begin to increase in intensity, indicating a commensurate component of the magnetic order is developing. Then at $T_{M3}$=18 K there are abrupt changes in the (600) and (020) intensities, where the heat capacity begins to increase together with a strong anomaly in the dielectric constant. A breaking of the structural inversion symmetry at 18 K has been reported from optical second-harmonic generation measurements[22], but this structural distortion could not be identified from our high-resolution neutron and synchrotron x-ray powder diffraction patterns, indicating that the distortion must be relatively small even though there is a noticeable increase in **P**. We note, however, that the present single crystal measurements are much more sensitive compared to powder diffraction, and thus it is possible that there is a small structural contribution to these intensity changes. Nevertheless, we conclude that the abrupt changes in the (600) and (020) peaks at 18 K are primarily magnetic in origin. We remark that magnetic scattering has been reported in a powder, but at a somewhat lower temperature,[26] and the orientation of **P** has been suggested to be in the *ac*-plane, which can be substantially enhanced by an applied magnetic field.[22]

The magnetic incommensurate wave vector varies regularly with temperature, with $q_h$ ranging from 0.355 to 0.417 and $q_k$ varying from zero to 0.159 (Fig. 3(c)). In addition, there is a peak in intensity exhibited at each transition (Fig. 3(d)). It is interesting to note that during the development of the electrical polarization, the ICM wave vector in the *ab*-plane is essentially temperature independent at $q_h$=0.415 and $q_k$=0 and the intensity becomes relatively weak but visible. Note that $q_h$ takes on its maximum value while the intensities of the ICM reflections are weak in the transition regime of electrical polarization. We also see that the temperature dependence of the intensity exhibits an identifiable peak in each of the three phase regimes between $T_{M1}$ and $T_{M4}$. The overall behavior of the ICM intensities and wave vector indicate that the magnetic degree-of-freedom is delicately balanced and strongly coupled to the electric degree-of-freedom.

The ICM wave vector also varies with applied magnetic field as shown in Fig. 4. Figures 4(a) and 4(b) show intensity maps of the ICM peaks, taken at 1.5 K, for B=0 and B=14 T applied along the *c*-axis direction. At 14 T we see that both $q_h$ and $q_k$ have shifted to substantially larger values (away from the structural Bragg peak), indicating a sizeable field dependence. The change in $q_h$ occurs abruptly while the magnitude above and below the transition is the same within experimental uncertainties as shown in Fig. 4(c). On the other hand, the critical field for triggering the abrupt change in $\vec{q}$ is temperature dependent, with $B_C$=10 T at 1.5 K and decreasing to 9 T at 8 K. Moreover, the intensity of the ICM reflection increases by 82% at 1.5 K through the transition, while that of the commensurate (020) Bragg reflection decreases by 78% as shown in Fig. 4(d). Similar behavior has also been observed for the ICM reflections associated with the (310) reflection. It has been reported that the effects on **P** also depend on the orientation of **B**, with a magnitude of **B** exceeding ≈9 T along the *c*-axis drastically suppressing **P**,[22] consistent with the data in Fig. 4. These observations show the direct interplay between the magnetic and electric order parameters.

In summary, we have observed a series of magnetic phase transitions as a function of temperature and magnetic field, that involve both commensurate and incommensurate components to the magnetic structures. Two of these transitions have a clear and direct relationship to the electric susceptibility, indicative of the strong coupling between the magnetic order and ferroelectricity. It will be interesting to determine the detailed magnetic structures in these various magnetically ordered phases, but such determinations will require rather extensive high resolution powder and single crystal diffraction measurements due to the large number of Co spins in the chemical unit cell.

**Acknowledgements.** We gratefully thank Brooks Harris, Qingzhen Huang and Taner Yildirim for fruitful discussions. This work was supported in part by the National Science Council of Taiwan under Grant Nos. NSC 98-2112-M-008-016-MY3 and NSC 100-2112-M-110-004-MY3.

**References**

[1]R. E. Cohen, Nature **358**, 136-139 (1992).

[2]H. Katsura, N. Nagaosa, and A. V. Balatsky, Phys. Rev. Lett. **95**, 057205 (2005).

[3]T. Masuda, A. Zheludev, B. Roessli, A. Bush, M. Markina, and A. Vasiliev, Phys. Rev. B **72**, 014405 (2005).

[4]G. Lawes, A. B. Harris, T. Kimura, N. Rogado, R. J. Cava, A. Aharony, O. Entin-Wohlman, T. Yildirim, M. Kenzelmann and C. Broholm, Phys. Rev. Lett. **95**, 087205 (2005).

[5]G. Lautenschlager, H. Weitzel, T. Vogt, R. Hock, A. Bohm, M. Bonnet and H. Fuess, Phys. Rev. B **48**, 6087-6098 (1993).

[6]O. Heyer, N. Hollmann, I. Klassen, S. Jodlauk, L. Bohaty, P. Becker, J. A. Mydosh, T. Lorenz and D. Khomskii, J. Phys.: Condens. Matter **18**, L471-L475 (2006).

[7]N. A. Hill, J. Phys. Chem. B **104**, 6694 (2000).

[8]D. Khomskii, Physics **2**, 20 (2009).

[9]S.-W. Cheong and M. Mostovoy, Nature Mater. **6**, 13-20 (2007). A. B. Harris, Phys. Rev. B **76**, 054447 (2007).

[10]O. P. Vajk, M. Kenzelmann, J. W. Lynn, S. B. Kim, and S.-W. Cheong, Phys. Rev. Lett. **94**, 087601 (2005).

[11]D. Lebeugle, D. Colson, A. Forget, M. Viret, P. Bonville, J. F. Marucco and S. Fusil, Phys. Rev. B **76**, 024116 (2007).

[12]P. Fischer, M. Polomska, I. Sosnowska and M. Szymanski, J. Phys. C: Solid St. Phys. **13**, 1931-1940 (1980).

[13]R. Seshadri and N. A. Hill, Chem. Mater. **13**, 2892-2899 (2001).

[14]G. R. Blake, L. C. Chapon, P. G. Radaelli, S. Park, N. Hur, S-W. Cheong and J. Rodríguez-Carvajal, Phys. Rev. B **71**, 214402 (2005).

[15]C. R. dela Cruz, F. Yen, B. Lorenz, M. M. Gospodinov, C. W. Chu, W. Ratcliff, J. W. Lynn, S. Park and S.-W. Cheong, Phys. Rev. B **73**, 100406(R) (2006).

[16]A. Munoz, Inorg. Chem. **40**, 1020-1028 (2001).

[17]B. Lorenz, Y. Q. Wang, Y. Y. Sun and C. W. Chu, Phys. Rev. B **70**, 212412 (2004).

[18]I. Živković, K. Prša, O. Zaharko and H. Berger, J. Phys.: Condens. Matter **22**, 056002 (2010).

[19]M. Herak, H. Berger, M. Prester, M. Miljak, I. Živković, O. Milat, D. Drobac, S. Popović and O. Zaharko, J. Phys.: Condens. Matter **17**, 7667-7679 (2010).

[20]S. A. Ivanov, P. Nordblad, R. Mathieu, R. Tellgren, C. Ritter, N. V. Golubko, E. D. Politova and M. Weil, Mater. Res. Bull. **46**, 1870 (2011).

[21]K. M. Rabe, C. H. Ahn and J.-M. Triscone, *Physics of ferroelectrics: a modern perspective* (Springer-Verlag, Berlin, 2007), p.17.

[22]M. Hudl, R. Mathieu, I. Ivanov, M. Weli, V. Carolus, Th. Lottermoser, M. Fiebig, Y. Tokunaga, Y. Tokura and P. Nordblad, Phys. Rev. B **84**, 180404(R), (2011).

[23]J. L. Her, C. C. Chou, Y. H. Matsuda, K. Kindo, H. Berger, K. F. Tseng, C. W. Wang, W. H. Li, and H. D. Yang, Phys. Rev. B **84**, 235123 (2011).

[24]R. Becker, M. Johnssona and H. Berger, Acta Cryst. C**62**, i67-i69 (2006).

[25]J. W. Lynn, Y. Chen, S. Chang, Y. Zhao, S. Chi, W. Ratcliff, B. G. Ueland, and R. W. Erwin, J. Research NIST **117**, 61 (2012).

[26]S. A. Ivanov, R. Tellgren, C. Ritter, P. Nordblad, R. Mathieu, G. Andre, N. V. Golubko, E. D. Politova, and M. Weil, Mater. Res. Bull. **47**, 63 (2012).

[27]A. B. Harris, Phys. Rev. B **85**, 100403(R) (2012).

FIG. 1. (Color online) Effects of applied magnetic field on the bulk susceptibility $\chi(T)$. Two critical points, designated $T_{M1}$ and $T_{M3}$ as described below, can be identified. An applied magnetic field separates the anomalies, reflecting that they are magnetic in origin. They indicate the initial development of magnetic order at $T_{M1}$, and the appearance of a magnetic transition at $T_{M3}$. Curves are just guides to the eye. (Note: 1 emu/(g Oe) = $4\pi\times10^{-3}$ $m^3$/kg).

FIG. 2 (Color online) Neutron diffraction measurements of scattering in the (*h*, *k*, 0) scattering plane, obtained using the PSD on BT-7.[25] (a)-(b) Intensity map observed at 22 K. The nuclear Bragg peaks at integer positions are accompanied by four satellite magnetic reflections, indicating the development of incommensurate (ICM) magnetic order. Note that the ordering wave vector is incommensurate in both *h* and *k*. There is also clear diffuse scattering surrounding the ICM peaks observed at this temperature,

which originates from inelastic magnetic scattering. (c) At 16.5 K the magnetic structure is incommensurate along the *h* direction, while the incommensurability along *k* vanishes, demonstrating that a magnetic phase transition has occurred. (d) At the base temperature of 2 K the incommensurability along the k direction is re-established, demonstrating that another phase transition has occurred. (e)-(g) Maps of the temperature dependence of three representative incommensurate magnetic peaks, showing how the modulation wave vectors and intensities vary with temperature.

FIG. 3. (Color online) Magnetic intensities and wave vectors as a function of temperature, together with the specific heat and dielectric constant. (a) Thermal variations of the (600) and (020) Bragg intensities. Above $T_{M1}$=26 K these peaks are purely nuclear, and there is no change in intensity, other than some magnetic critical scattering around $T_{M1}$, until the intensity starts to increase at $T_{M2}$, revealing that a commensurate component to the magnetic scattering has developed. A small change in the nuclear (structural) intensity also cannot be ruled out. (b) Temperature dependence of the heat capacity and dielectric constant. It is interesting to note that the clear transitions associated with the commensurate magnetic order parameter at $T_{M2}$ and $T_{M3}$ are in good agreement with changes in behavior of the electrical polarization, demonstrating the direct interplay between the magnetic and electric degrees-of-freedom. (c)-(d) Temperature variation of both components of the incommensurate wave vector, and integrated intensity, of the ICM magnetic order. Curves connecting data points are guides to the eye.

FIG. 4. (Color online) Magnetic field dependence of the commensurate and incommensurate magnetic order. (a)-(b) Intensity maps around the (020) Bragg reflection and its ICM counterparts at 1.5 K, taken with (a) B=0 and (b) 14 T applied along the *c*-axis. The elongated appearance of the peaks is due to the quite different *x* and *y* scales for the plot; the peaks are resolution limited in both directions. Note that both components of the incommensurate wave vector are larger at high field. (c) ICM wave vector components as a function of field. At 1.5 K there is an abrupt transition observed at ≈10 T, which shifts to ≈9 T at 8 K. (d) The high field transition is also reflected in the intensities of both the commensurate and incommensurate magnetic peaks. Curves connecting data points are guides to the eye.

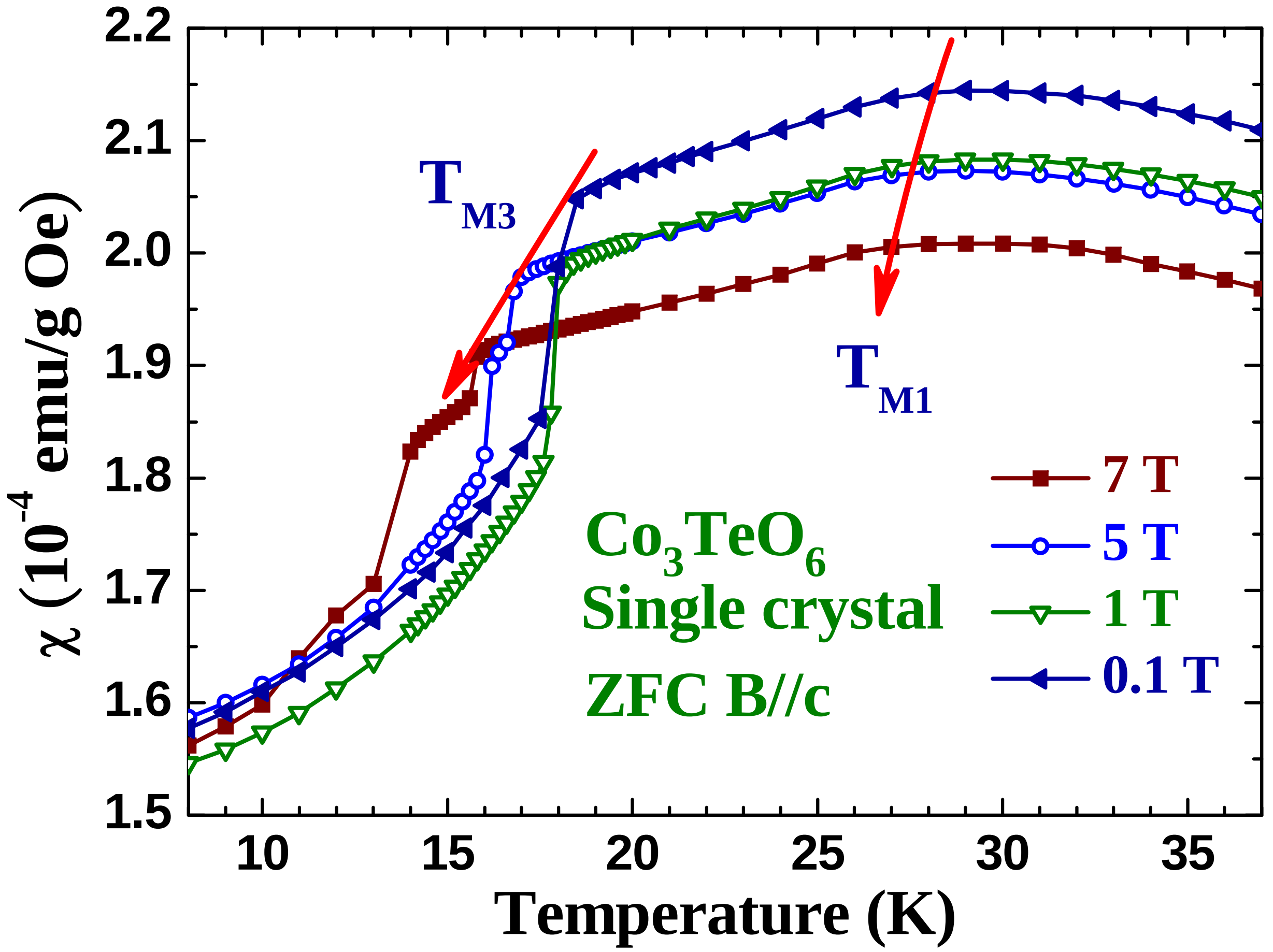

Fig. 1

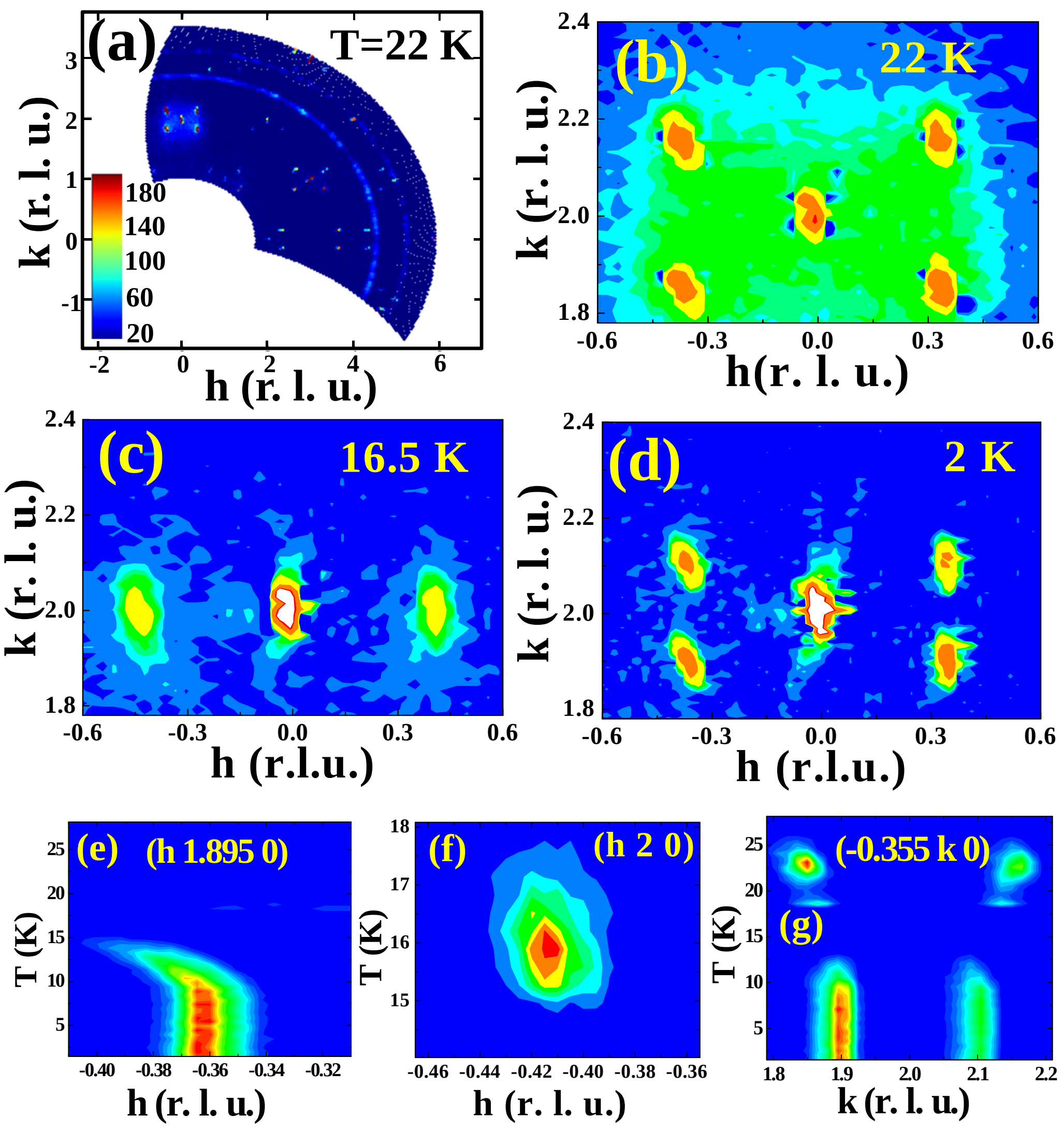

(a)
T=22 K
k (r. l. u.)
h (r. l. u.)
180
140
100
60
20
(b)
22 K
k (r. l. u.)
h(r. l. u.)
(c)
16.5 K
k (r. l. u.)
h (r.l.u.)
(d)
2 K
k (r. l. u.)
h (r.l.u.)
(e)
(h 1.895 0)
T (K)
h (r. l. u.)
(f)
(h 2 0)
T (K)
h (r. l. u.)
(-0.355 k 0)
(g)
T (K)
k (r. l. u.)

Fig. 2

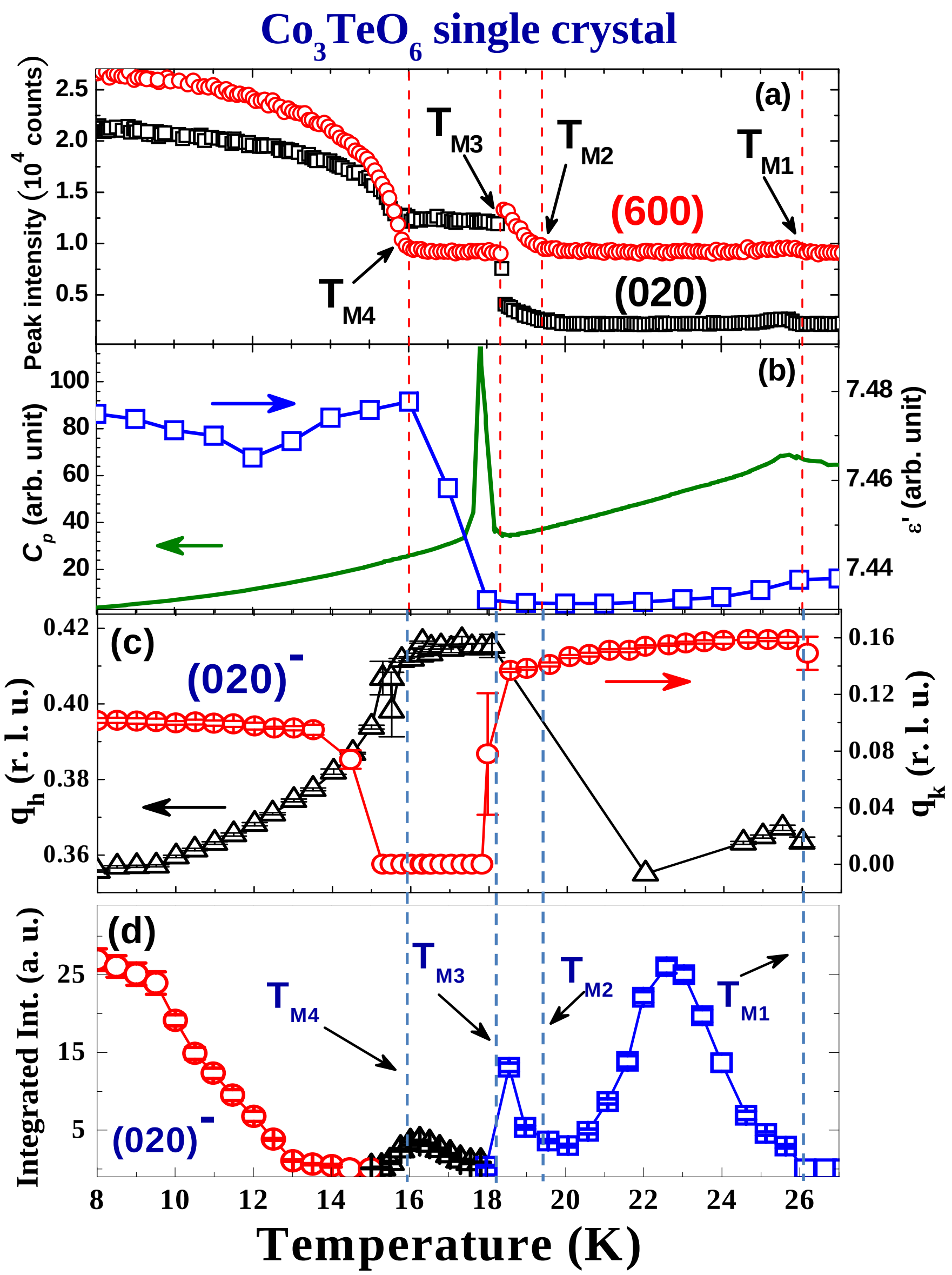

Fig. 3

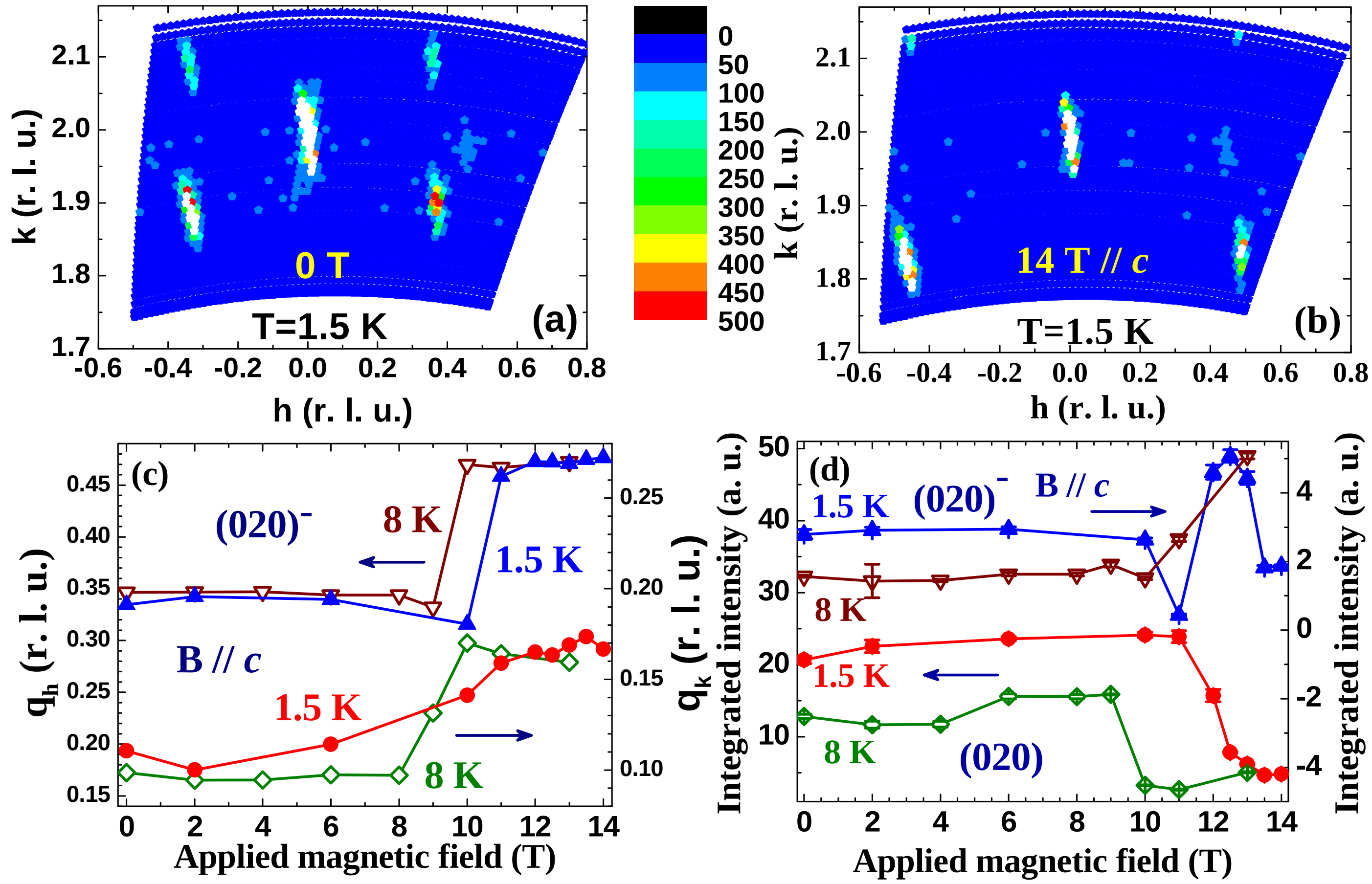

0 T
T=1.5 K
(a)
k (r. l. u.)
h (r. l. u.)
14 T // c
T=1.5 K
(b)
(c)
(020)⁻
8 K
1.5 K
B // c
1.5 K
8 K
q_h (r. l. u.)
Applied magnetic field (T)
(d)
1.5 K
(020)⁻
B // c
8 K
1.5 K
8 K
(020)
q_k (r. l. u.) Integrated intensity (a. u.)
Integrated intensity (a. u.)
Applied magnetic field (T)

Fig. 4